\documentclass[a4paper,12pt]{article}
\usepackage{epsfig}
\begin{document}

\begin{center}

{\large\bf Particle-hole symmetry breaking due to Pauli blocking } \bigskip
 
 \footnotesize

D. Bonatsos$^1$, I. E. Assimakis$^1$, A. Martinou $^1$,  S. Sarantopoulou$^1$, 
S. Peroulis $^2$, N. Minkov$^3$

\medskip

{\it 
$^1$ Institute of Nuclear and Particle Physics, National Center for Scientific Research  ``Demokritos'', Athens, Greece 

$^2$ University of Athens, Faculty of Physics, Zografou Campus, GR-15784 Athens, Greece

$^3$ Institute of Nuclear Research and Nuclear Energy, Bulgarian Academy of Sciences, 
72 Tzarigrad Road, 1784 Sofia, Bulgaria

} 

\end{center}

\rule{16.5cm}{0.3mm}

\bigskip

\footnotesize
{\bf Abstract} \hskip 5mm  
Particle-hole symmetry has been used on several occasions in nuclear structure over the years. We prove that particle-hole symmetry is broken in nuclear shells possessing the proxy-SU(3) symmetry. The breaking of the symmetry is rooted in the Pauli principle and the short range nature of the nucleon-nucleon interaction. The breaking of the symmetry explains the dominance of prolate over oblate shapes in deformed nuclei and determines the regions of prolate to oblate shape transitions in the nuclear chart. Furthermore, it is related to the existence of specific regions of shape coexistence across the nuclear chart, surrounded by regions in which shape coexistence does not occur.  

\bigskip


 \rule{16.5cm}{0.3mm}
\pagestyle{empty}

\normalsize

\bigskip\noindent
{\bf INTRODUCTION}

\medskip\noindent

Particle-hole symmetry in nuclear shells has been used over the years in a variety of theoretical approaches, as, for example, in counting the bosons describing a nucleus in the framework of the Interacting Boson Model \cite{IA}, or in the construction of the $N_p N_n$ scheme and the $P$-factor \cite{Casten}, which allow for the determination of properties of unknown nuclei by interpolation instead of extrapolation. 
On several occasions it has turned out to be a successful approximation. However, by looking at the diagram of orbitals in the shell model \cite{Ring}, or at the Nilsson diagrams \cite{Nilsson1,Nilsson2}, which correspond to a simplified version of the shell model based on the spin-orbit interaction and a 3-dimensional harmonic oscillator (3D-HO) with cylindrical symmetry, allowing the model to take into account the tendency of nuclei away from closed shells to acquire deformed prolate or oblate shapes \cite{BM}, one easily sees that the orbitals near the bottom and near the top of a given nuclear shell are characterized by very different values of the relevant quantum numbers, indicating that, for certain purposes at least, the use of particle-hole symmetry should be avoided. 

On the other hand, a new approximate symmetry appropriate for heavy nuclei has been recently introduced,
called the proxy-SU(3) symmetry \cite{proxy1,proxy2,proxy3}. It will be shown that particle-hole symmetry breaking arises naturally 
within this scheme, based solely on group theoretical mathematical rules, the Pauli principle, and the short range nature of the nucleon-nucleon interaction \cite{Casten,Ring}. Furthermore, it will be demonstrated that 
the particle-hole symmetry breaking within nuclear shells is responsible for the prolate over oblate 
dominance \cite{HM} in deformed nuclei, as well as for the prolate to oblate shape/phase transition 
\cite{Linnemann} observed in the heavy rare earths. In addition, the role of particle-hole symmetry breaking in the creation of regions of shape coexistence \cite{odd,Wood,Heyde} in specific places within the nuclear chart will be demonstrated. 

\bigskip\noindent
{\bf THE PROXY-SU(3) SCHEME}
\medskip\noindent

The proxy-SU(3) symmetry is based on the fact that Nilsson orbitals with similar values of quantum numbers 
possess large spatial overlaps, thus maximizing the interaction among them. This idea dates back to the seventies,
when Federman and Pittel \cite{FP1,FP2} demonstrated that proton-neutron pairs of orbitals with similar values of quantum numbers play a crucial role in the onset of nuclear deformation. More recently, the importance of 
proton-neutron pairs belonging to Nilsson orbitals characterized by similar sets of quantum numbers has been realized within a detailed study of double differences of nuclear masses \cite{Cakirli1,Cakirli2}.  
In this case, Nilsson orbitals characterized by quantum numbers differing by $\Delta K[\Delta N \Delta n_z \Delta\Lambda]=0[110]$ have been used, where $\Lambda$, $K$, $\Sigma$ are the projections of the orbital angular momentum, total angular momentum, and spin on the $z$-axis, while $N$ and $z$ are the total number of quanta and the number of quanta in the $z$-axis respectively. It has been proved \cite{Karampagia} that these 0[110] pairs of orbitals possess large spatial overlaps, since they have the same orientation in space.

Proxy-SU(3) occurred  by realizing that the same mathematical reasons leading to large spatial overlaps 
and therefore large interactions among 0[110] proton-neutron pairs, also lead to large spatial overlaps and therefore large interactions for proton-proton-pairs and neutron-neutron pairs. This fact leads to an approximate solution of the long standing problem of restoration of the SU(3) symmetry of the isotropic three-dimensional harmonic oscillator (3D-HO) \cite{Wybourne} in medium mass and heavy nuclei beyond the sd shell.
While the 3D-HO SU(3) symmetry is known to hold up to the sd shell since the path-breaking work of Elliott in 1958 \cite{Elliott1,Elliott2,Elliott3}, it is well known that the SU(3) symmetry is broken beyond the sd shell because of the strong spin-orbit interaction, resulting in each shell in the loss of a number of orbitals,
which escape to the shell below, and in parallel in the gain of another bunch of orbitals coming down from the shell above \cite{Ring}. The new set of orbitals now occupying the shell does not possess the SU(3) symmetry. But the SU(3) symmetry can be approximately restored by noticing that the invading from above orbitals are 
0[110] counterparts of the orbitals which escaped to the shell below, except one invading orbital which has no counterpart. In this way the SU(3) symmetry is restored, but it encompasses all the orbitals in the new shell but one. As a consequence, the proxy-SU(3) shells can accommodate two particles less than the 
corresponding original shells of the shell model. For example, the sdg shell in proxy SU(3) can accommodate
30 like nucleons, while the corresponding shell of the shell model is the 50-82 shell, which can accommodate 32 like nucleons. Detailed calculations using the original Nilsson shells and the proxy-SU(3) shells have been reported in Ref. \cite{proxy1}, revealing that the quality of the approximation improves as one moves to higher shells. 

In the subsequent sections, some of the consequences of the proxy-SU(3) symmetry will be reviewed.

%%%%%%%%%%%%%%%%%%%%%%%% Table 1  %%%%%%%%%%%%%%%%%%%%%%%%%%%%%%%
\begin{table}[htb]

\caption{\footnotesize{Highest weight SU(3) irreps for U(n), n=6, 10, 15, 21, 28, 36. Irreps breaking the particle-hole symmetry are indicated by boldface. }}
\smallskip
\small\noindent\tabcolsep=9pt
\begin{center}
\footnotesize{
\begin{tabular}{ r l r r r r r r } 
\hline
\hline
\\[-8pt]

   &            & 8-20 & 28-50 & 50-82     & 82-126     &126-184&184-258\\
   &            & sd   & pf    & sdg       &  pfh       & sdgi  & pfhj  \\
M  & irrep      & U(6) & U(10) & U(15)     & U(21)      & U(28) & U(36) \\
 0 &            &(0,0) &(0,0)  &(0,0)      &(0,0)       & (0,0) & (0,0) \\  
 2 & [2]        &(4,0) & (6,0) & (8,0)     &(10,0)      &(12,0) & (14,0)\\ 
 4 & [$2^2$]    &(4,2) & (8,2) &(12,2)     &(16,2)      &(20,2) & (24,2)\\
 6 & [$2^3$]    &(6,0) &(12,0) &(18,0)     &(24,0)      &(30,0) & (36,0)\\
 8 & [$2^4$]    &(2,4) &(10,4) &(18,4)     &(26,4)      &(34,4) & (42,4)\\
10 & [$2^5$]    &(0,4) &(10,4) &(20,4)     &(30,4)      &(40,4) & (50,4)\\
12 & [$2^6$]    &(0,0) &{\bf(12,0)}&(24,0) &(36,0)      &(48,0) & (60,0)\\
14 & [$2^7$]    &      &{\bf(6,6)} &(20,6) &(34,6)      &(48,6) &(62,6) \\
16 & [$2^8$]    &      & (2,8) &{\bf(18,8)}&(34,8)      &(50,8) &(66,8)  \\
18 & [$2^9$]    &      & (0,6) &{\bf(18,6)}&(36,6)      &(54,6) &(72,6) \\
20 & [$2^{10}$] &      & (0,0) &{\bf(20,0)}&(40,0)      &(60,0) &(80,0) \\
22 & [$2^{11}$] &      &       &{\bf(12,8)}&{\bf(34,8)} &(56,8) &(78,8) \\
24 & [$2^{12}$] &      &       &{\bf(6,12)}&{\bf(30,12)}&(54,12) &(78,12) \\
26 & [$2^{13}$] &      &       &(2,12)     &{\bf(28,12)}&(54,12) &(80,12) \\
28 & [$2^{14}$] &      &       & (0,8)     &{\bf(28,8)} &(56,8) &(84,8) \\
30 & [$2^{15}$] &      &       & (0,0)     &{\bf(30,0)} &{\bf(60,0)}&(90,0) \\
32 & [$2^{16}$] &      &       &           &{\bf(20,10)}&{\bf(52,10)} &(84,10)\\
34 & [$2^{17}$] &      &       &           &{\bf(12,16)}&{\bf(46,16)} &(80,16)\\
36 & [$2^{18}$] &      &       &           &{\bf(6,18)} &{\bf(42,18)} &(78,18) \\
38 & [$2^{19}$] &      &       &           &(2,16)      &\bf{(40,16)} &\bf{(78,16)} \\
40 & [$2^{20}$] &      &       &           &(0,10)      &\bf{(40,10)} &\bf{(80,10)} \\
42 & [$2^{21}$] &      &       &           &(0,0)       & \bf{(42,0)} &\bf{(84,0)} \\

\hline
\end{tabular}   }
\end{center}

\end{table}

\bigskip\noindent
{\bf BREAKING OF THE PARTICLE-HOLE SYMMETRY}
\medskip\noindent

In order to discuss this topic, some group theoretical language should be introduced first. In particular, 
the $n$-th shell of the 3D-HO has a unitary symmetry U($(n+1)(n+2)/2$) \cite{Wybourne}, possessing  an SU(3) subalgebra \cite{BK}. The possible states of a number of nucleons within such a shell are classified according to the irreducible representations (irreps) of SU(3) occurring for this number of particles. 
For SU(3) irreps, the Elliott \cite{Elliott1,Elliott2} notation $(\lambda,\mu)$  is used, where $\lambda$ 
and $\mu$ are the Elliott quantum numbers. The Elliott quantum numbers are related to the numbers $f_1$ and $f_2$, indicating the numbers of boxes in the first row and in the second row, respectively, of the relevant Young diagram, through the relations $\lambda=f_1-f_2$, $\mu=f_2$. The list of SU(3) irreps appearing in a given U($n$) for each number of protons or neutrons is provided by existing codes \cite{code}. The irreps in these lists do not appear in a random way. The highest weight (h.w.) irrep appears on the top, followed by the other irreps in order of decreasing weight. Loosely speaking, highest weight means highest probability for this irrep to appear in a physical system. Lists of h.w. irreps for the U($n$) of interest are given in tables in Refs. \cite{proxy2,37J,40GC} for both even and odd numbers of particles. A partial list of irreps 
for even number of particles is given here in Table 1. 

Looking at these lists, one immediately sees that particle-hole symmetry is broken. The irreps beyond midshell which break the particle-hole symmetry appear in the tables in boldface. We shall demonstrate this through an example. 

Let us consider 6 particles in the proxy-SU(3) pfh shell, having the U(21) symmetry. From Table 1 we see that the corresponding h.w. irrep is (24,0). Now let us look at 6 holes in this shell. These correspond to 
$42-6=36$ particles, for which the relevant irrep from Table 1 is (6,18). If particle-hole symmetry were present, the irrep for 6 holes would have been the conjugate of the irrep of 6 particles, i.e., the conjugate of (24,0), which is (0,24). 

Let us now try to understand the physical origins of this result, bearing in mind that the nucleon-nucleon force 
has a short range \cite{Casten,Ring}, therefore favoring the most symmetric spatial configurations, i.e., the most symmetric irreps for the spatial part of the nuclear wave function, since these are the configurations which will provide maximal spatial overlaps of the wave functions. In parallel, we should remember that the Pauli principle is present, therefore the total wave function has to be fully antisymmetric. As a result, the spin-isospin part of the total wave function tends to be as antisymmetric as possible. In the case of two particles, we know that this leads to the well known result that for  total spin $S$ and total isospin $T$, the ($S$, $T$) pairs preferred by nature are (1,0) and (0,1). When many particles are considered, the spin-isospin part becomes as antisymmetric as possible, while the spatial part becomes as symmetric as possible. In group theoretical language, the irrep corresponding to the spin-isospin part will be the conjugate of the irrep corresponding to the spatial part, so that overall antisymmetry 
will be guaranteed. 

Coming back to our example, let us compare the irreps (6,18) and (0,24). The former corresponds to a Young diagram with 24 boxes in its first row and 18 boxes in its second row, while the Young diagram for the latter has 24 boxes in the first row and 24 boxes in the second row. Since we know by definition in the Young diagrams that boxes in the same row indicate symmetrization, while boxes in the same column mean antisymmetrization, we realize that (0,24) contains more antisymmetrizations than (6,18), therefore 
(6,18) is more symmetric than (0,24). We conclude that the breaking of the particle-hole symmetry is a consequence of the Pauli principle and the short range of the nucleon-nucleon interaction. 

In other words, in the first half of the shell the fermions are not crowded too much, therefore the limitations imposed by the Pauli principle can be easily accommodated. In contrast, beyond midshell the fermions get crowded, thus the Pauli principle restrictions affect also the spatial part, in order to guarantee the overall antisymmetry of the full wave function. In fact, the h.w. spatial irreps, seen in Table 1, are the most symmetric spatial irreps  allowed by the Pauli principle. The full proof of this statement requires use of the Gelfand-Tsetlin diagrams \cite{code}. A detailed discussion of this proof can be found in Ref. \cite{MartinouPhD}. 

It is worth noticing that the mathematical result of particle-hole symmetry breaking has been known since many years. It already exists in the first article  by Elliott \cite{Elliott1} on the sd shell, although in this case it can be observed only for 7 particles. It can also be seen in Table 5 of  Ref. \cite{pseudo1},
in which the h.w. irreps in the U(10) shell for odd numbers of particles are reported. However, its important consequences appear to have escaped attention since then.

 \bigskip\noindent
{\bf PROLATE OVER OBLATE DOMINANCE AND PROLATE-OBLATE \\ SHAPE/PHASE TRANSITION }
\medskip\noindent 

The result of the previous section has many important consequences. First, it provides an answer to the long standing question \cite{HM} of prolate over oblate dominance in the ground states of even-even nuclei. 
Looking in detail into the rare earths with 50-82 protons and 82-126 neutrons, for example, one can use the h.w. irreps corresponding to their valence protons and neutrons and determine in this way the irrep corresponding to the ground state of each nucleus. The relevant table can be seen in Ref. \cite{proxy2}. 
One immediately sees that most irreps have $\lambda >\mu$, corresponding to prolate (rugby ball) shapes, 
while few of them have $\lambda <\mu$, corresponding to oblate (pancake) shapes. Similar tables result 
in other nuclear shells as well \cite{proxy2}.  

A second consequence of the particle-hole symmetry breaking is the existence of a shape/phase transition 
from prolate to oblate shapes \cite{Linnemann}. Looking once more at the rare earths shell with 50-82 protons and 82-126 neutrons, one sees that this transition occurs in nuclei located near the top of the proton shell and near the top of the neutron shell simultaneously. More details can be seen in the tables provided in Ref. \cite{proxy2}.

\bigskip\noindent
{\bf SHAPE COEXISTENCE}\label{coex}
\medskip\noindent

In this section we will show that shape coexistence can also be considered as a consequence of the particle-hole symmetry breaking. 

Shape coexistence \cite{odd,Wood,Heyde}
is a subtle nuclear phenomenon, said to occur in atomic nuclei in which two low-lying 
$0^+$ states, or whole bands based on them, appear very close in energy but possess clearly different structure and/or shapes. 
Providing a justification for shape coexistence is a formidable problem for nuclear theory \cite{Poves}, ranging from extensive numerical calculations in the framework of relativistic mean field \cite{Lalazissis}
to group theoretical approaches, either fermionic \cite{Elliott1,Elliott2} or bosonic \cite{IA}. 
From the experimental point of view, the observation of the energy levels of two bands being candidates 
for exhibiting shape coexistence is desirable to be completed by detailed studies of electromagnetic transition rates, $B(E0)$s and $B(E2)$s for example, demonstrating the substantial difference in properties and shapes of the members of the two bands.    

\bigskip\noindent
{\bf The $Z\sim 82$ region}
\medskip\noindent

A textbook example of shape coexistence is provided by the $_{82}$Pb, $_{80}$Hg and $_{78}$Pt isotopes, i.e. at or near the $Z=82$ magic number \cite{Wood,Heyde,Ramos}. 

One can start with the Pb ($Z=82$) isotopes, which have no valence protons in the shell model description.
Collected data for the $_{82}$Pb can be seen in Fig. 12 of the review article \cite{Heyde}, as well as 
in Fig. 1 of Ref. \cite{Ramos}. One sees that there are excited bands for $N=98$-108, with candidates existing also at $N=110$, 112,  which form a nearly parabolic intrusion with a local minimum at $N=104$. In addition, there are nearby-in-energy ground state bands starting from $N=104$, which extend up to $N=202$.  

One can next consider the Hg ($Z=80$) isotopes, for which data have been  collected in Fig. 10 of the review 
\cite{Heyde}, as well as in Fig. 2 of Ref. \cite{Ramos}. Again, one sees $K=0^+$ bands at $N=96$-110 forming a nearly parabolic intrusion. In these isotopes, one also sees nearby lying ground state bands all the way from $N=94$ to 126.

Data for the  Pt ($Z=78$) isotopes have been collected in Fig. 3.26 of the review \cite{Wood}, 
as well as in Fig. 3 of Ref. \cite{Ramos}. We see again  a parabolic intrusion by bands starting at $N=96$ and extending up to $N=110$, while all over from $N=90$ to 126 one also sees additional $K=0$ bands.
 
 \bigskip\noindent
{\bf The $Z\sim 50$ region}
\medskip\noindent

Another textbook example of shape coexistence is provided by the $_{50}$Sn isotopes, i.e. in the  region of the $Z=50$ magic number \cite{Wood,Heyde}. 

Data for shape coexistence in the Sn ($Z=50$) isotopes, which have no valence protons in the shell model description, have been collected in Fig. 3.10 of Ref. \cite{Wood}. We see a nearly parabolic intrusion starting from $N=60$ and extending up to $N=70$, with a local minimum at $N=66$. Additional $K^\pi=0^+$ bands appear all over from $N=56$ to 82, which are the ground state bands. 

In summary, we see that in all cases mentioned above there are low-lying bands forming a nearly parabolic intrusion, which  lie close in energy with the ground state bands. These intruder bands appear to be limited 
within the region $N=96$-110 near the $Z=82$ shell closure, while they appear to be confined in the region
$N=60$-70 near the $Z=50$ shell closure. 

\bigskip\noindent
{\bf Collective variables}
\medskip\noindent

We shall see below that these features of shape coexistence regions can be produced using symmetry arguments alone. The only quantity we are going to use is the second order Casimir operator of SU(3) \cite{IA}.
Using the Elliott \cite{Elliott1,Elliott2} notation $(\lambda, \mu)$ for the irreps of SU(3), the eigenvalues of the Casimir operatore read \cite{IA}
 \begin{equation}\label{C2} 
 C_2(\lambda,\mu)= \lambda^2+\lambda \mu + \mu^2+ 3\lambda +3 \mu. 
\end{equation}
It has been shown \cite{Castanos,Park} that the deformation variables $\beta$ and $\gamma$ of the collective model \cite{BM} can be expressed in terms of the Elliott quantum numbers $\lambda$ and $\mu$.
In particular, $\beta^2$ turns out to be proportional to $C_2(\lambda,\mu)+3$, while $\gamma$ is given by
\cite{Castanos,Park}
\begin{equation}\label{g1}
\gamma = \arctan \left( {\sqrt{3} (\mu+1) \over 2\lambda+\mu+3}  \right),
\end{equation}
which is parameter free. $\beta$ describes the magnitude of the deviation of the nuclear shape from sphericity, while $\gamma$ is related to triaxiality, with $\lambda>\mu$ corresponding to prolate (cigar-like) shapes, $\lambda<\mu$ describing oblate (pancake-like) shapes, and $\lambda=\mu$ providing maximal triaxiality.  
  
\bigskip\noindent
{\bf Competition between proxy-SU(3) and 3D-HO magic numbers}
\medskip\noindent

The basic idea behind shape coexistence appears to be the competition between spin-orbit magic numbers 
(14, 28, 50, 82, 126, 184, \dots), known to appear in the shell model, and three-dimensional harmonic oscillator (3D-HO) magic numbers (2, 8, 20, 40, 70, 112, 168, \dots), which hold in the absence of spin-orbit interaction. This competition has been pointed out in the review article of Ref.
\cite{Sorlin}, in which the evolution of nuclear magic numbers as one moves away from stability has been considered in detail. It is also known from relativistic mean field calculations \cite{Lalazissis}
that the spin-orbit force is reduced as one moves away from stability into the neutron-rich region. 

In medium-mass and heavy nuclei one can use the proxy-SU(3) scheme described in Refs. \cite{proxy1,proxy2}. Then to each nucleus corresponds a highest weight irrep, as described in Refs. \cite{proxy1,proxy2}, which is determined only by the number of valence protons and valence neutrons corresponding to this nucleus within the usual shell model (i.e., counting valence particles from the appropriate magic number of the shell model up). In parallel, one can determine the h.w. irrep which corresponds to the same nucleus
in the case in which the 3D-HO magic numbers prevail. In this case the valence particles 
will be counted from the relevant 3D-HO magic number up. The needed irreps can be read from tables 
given in Refs. \cite{proxy2,37J}. This will be clarified below through two examples.

a) Consider 64 nucleons. Within the proxy-SU(3) sdg shell, which starts at 50 nucleons and possesses the 
U(15) symmetry, these correspond to 14 valence particles, thus the relevant h.w. irrep is (20,6), in agreement with Table 1. If we consider the 3D harmonic oscillator magic numbers, the 64 nucleons correspond to 24 valence particles within the 40-70 shell, which possesses the U(15) symmetry,
thus from Table 1 we see that the relevant h.w. irrep is (6,12). In both cases 
the number of HO quanta is $N_q=4$, thus the Casimir eigenvalues of the two irreps are directly comparable. 

b) Consider 76 nucleons. Within the proxy-SU(3) sdg shell, which starts at 50 nucleons and possesses the 
U(15) symmetry, having $N_q=4$, these correspond to 26 valence particles, thus the relevant h.w. irrep is (2,12), in agreement with Table 1. If we consider the 3D harmonic oscillator magic numbers, the 76 nucleons correspond to 6 valence particles within the 70-112 shell, which possesses the U(21) symmetry, having $N_q=5$, 
thus from Table 1 we see that the relevant h.w. irrep is (24,0).
In this case the number of quanta is different in the two shells involved, thus one has to take this into account when trying to compare the Casimir eigenvalues of the two irreps.

%%%%%%%%%%%%%%%%%%%%%%%%%%% FIG.1 %%%%%%%%%%%%%%%%%%%%%%%%%%%%%%%%%%%%%%%%%%%

\begin{figure*}[htb]

{\includegraphics[width=90mm]{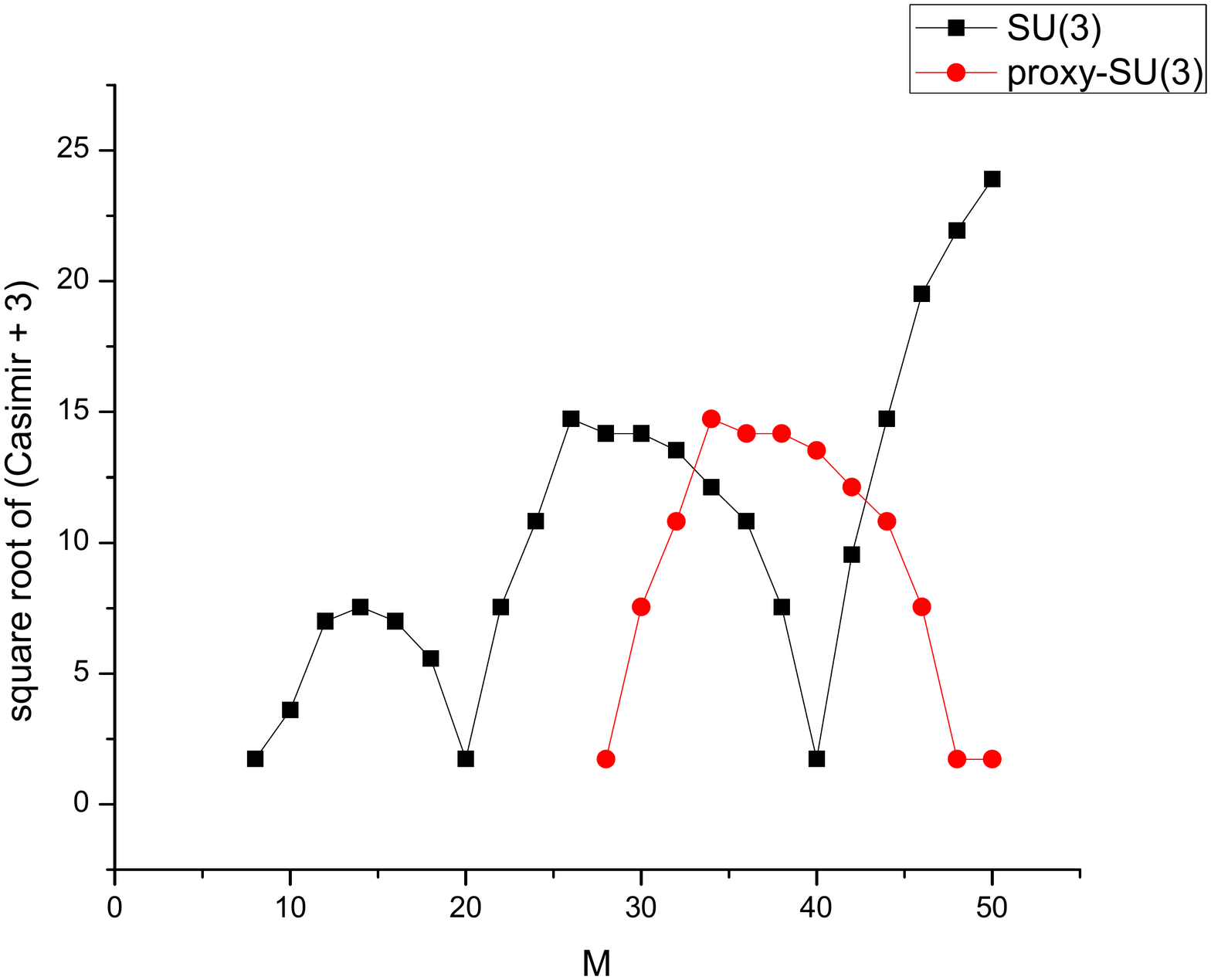}\hspace{1mm} 
\includegraphics[width=90mm]{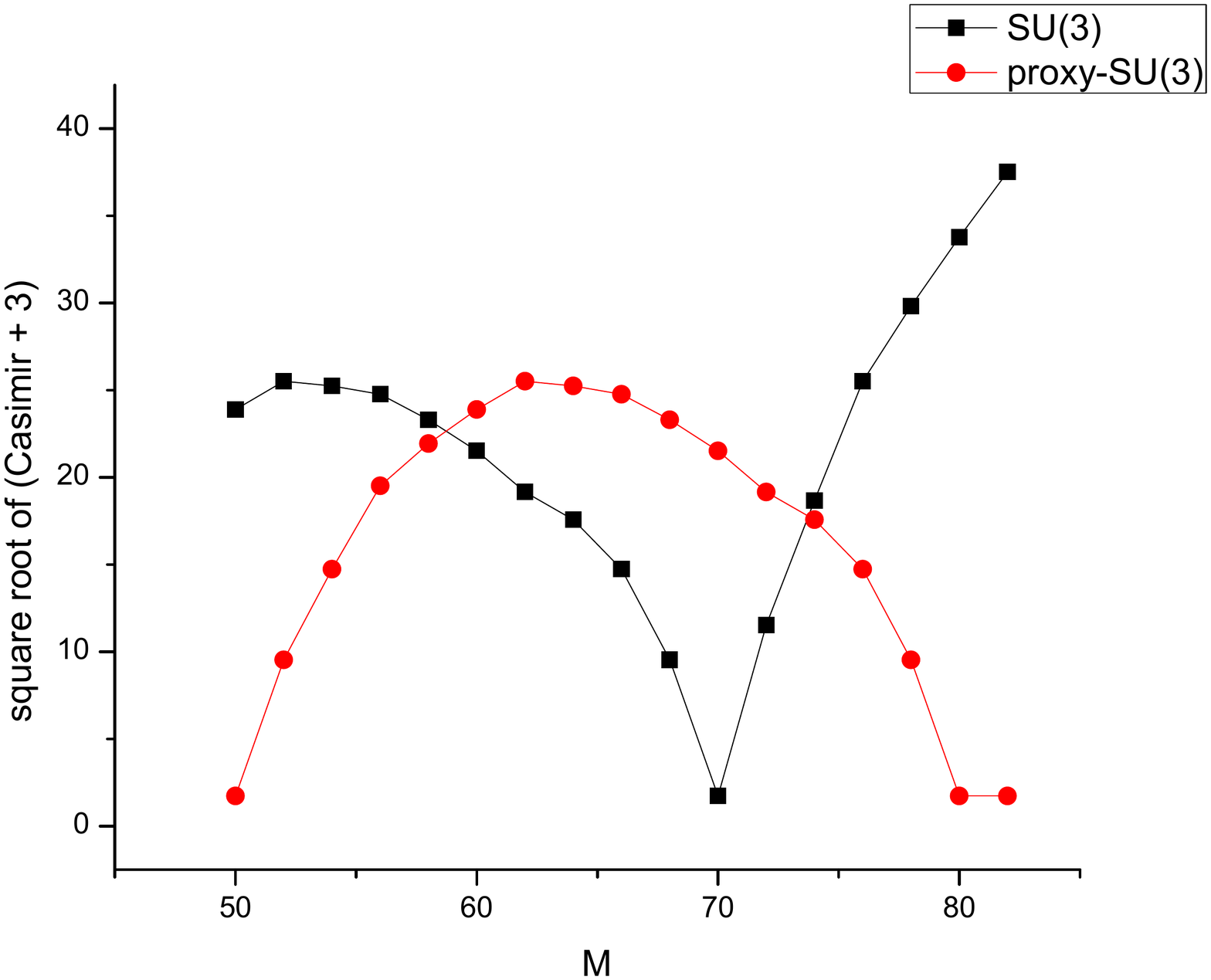}}
{\includegraphics[width=90mm]{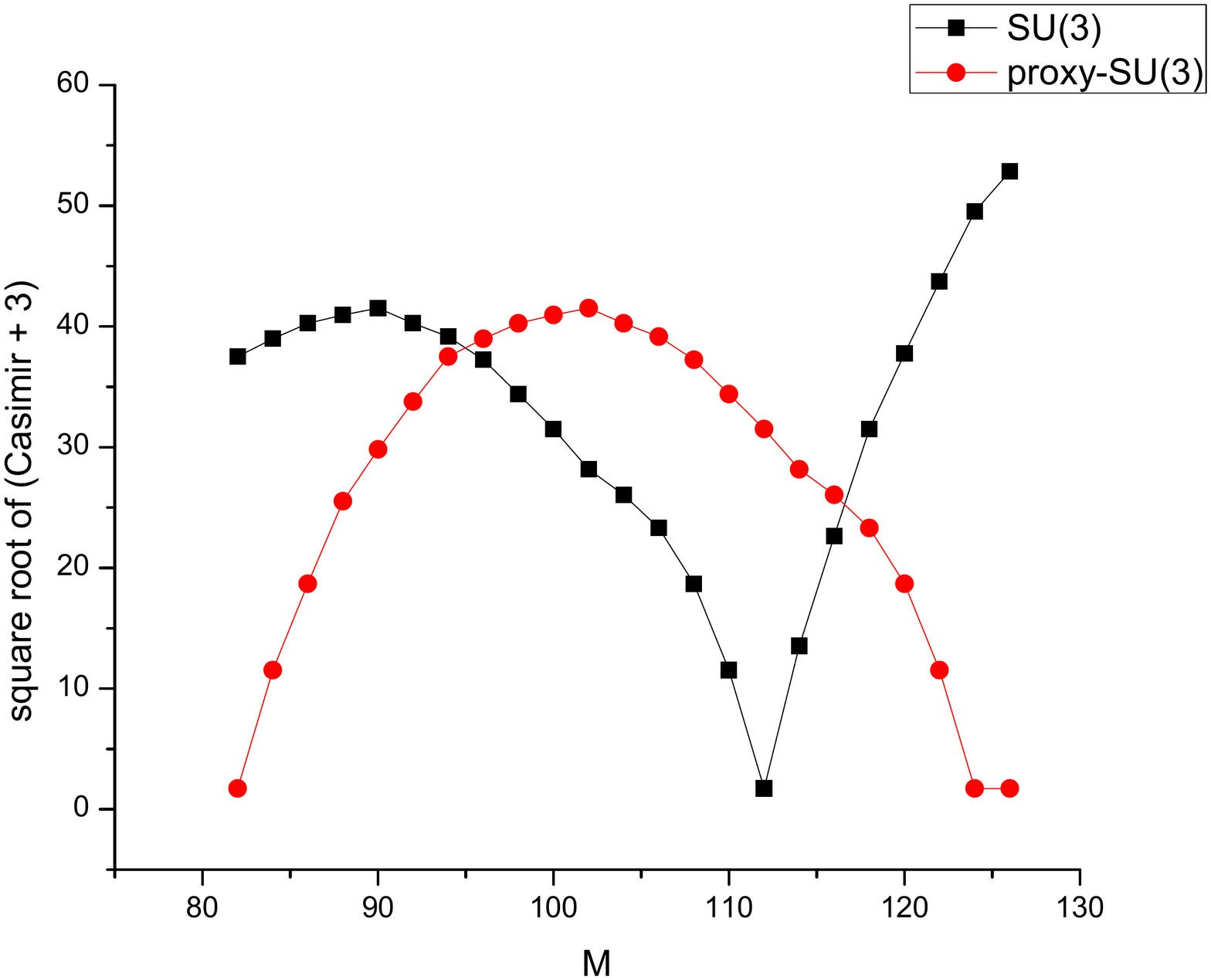}\hspace{1mm}
\includegraphics[width=90mm]{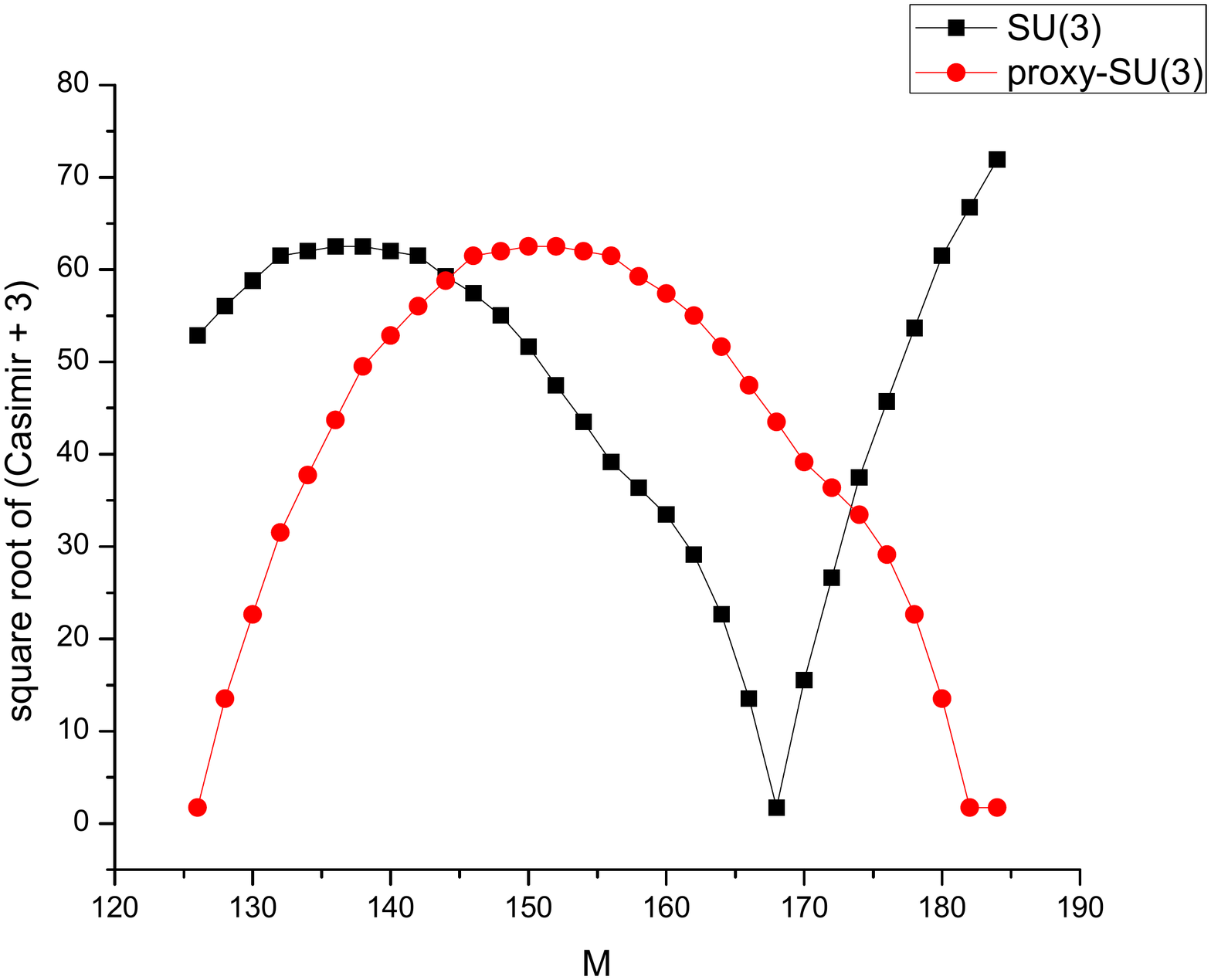}}

\caption{\footnotesize{The quantity  $\sqrt{C_2(\lambda,\mu)+3}$, which is roughly proportional to the deformation parameter $\beta$ \cite{Castanos,Park,proxy2}, is plotted against the nucleon (proton or neutron) number $M$ for the 3D harmonic oscillator possessing the SU(3) symmetry, as well as for the proxy-SU(3) scheme. See Section on Shape Coexistence for further discussion.}} 

\end{figure*}

%%%%%%%%%%%%%%%%%%%%%%%%%%% FIG.2 %%%%%%%%%%%%%%%%%%%%%%%%%%%%%%%%%%%%%%%%%%%

\begin{figure*}[htb]

{\includegraphics[width=90mm]{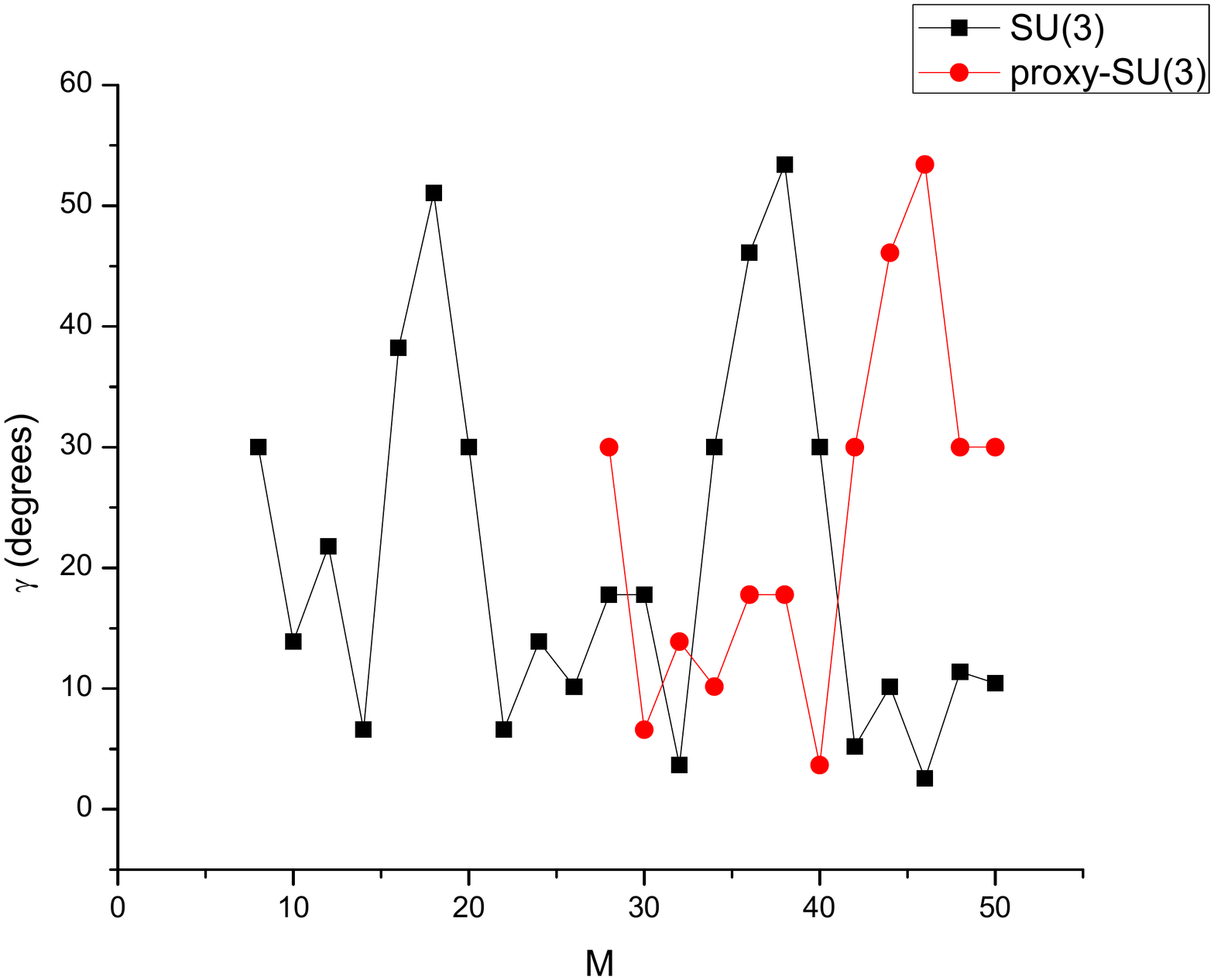}\hspace{1mm} 
\includegraphics[width=90mm]{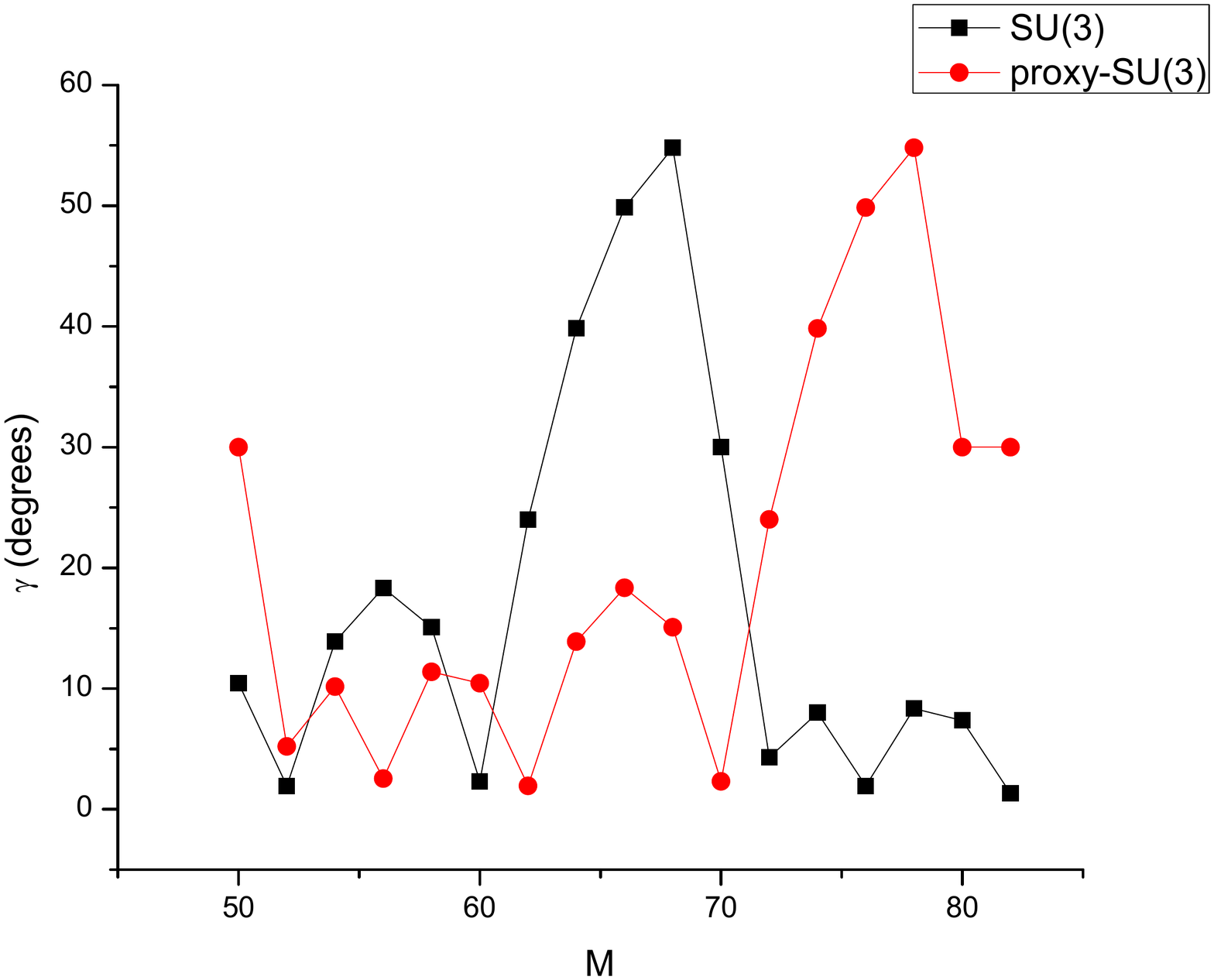}}
{\includegraphics[width=90mm]{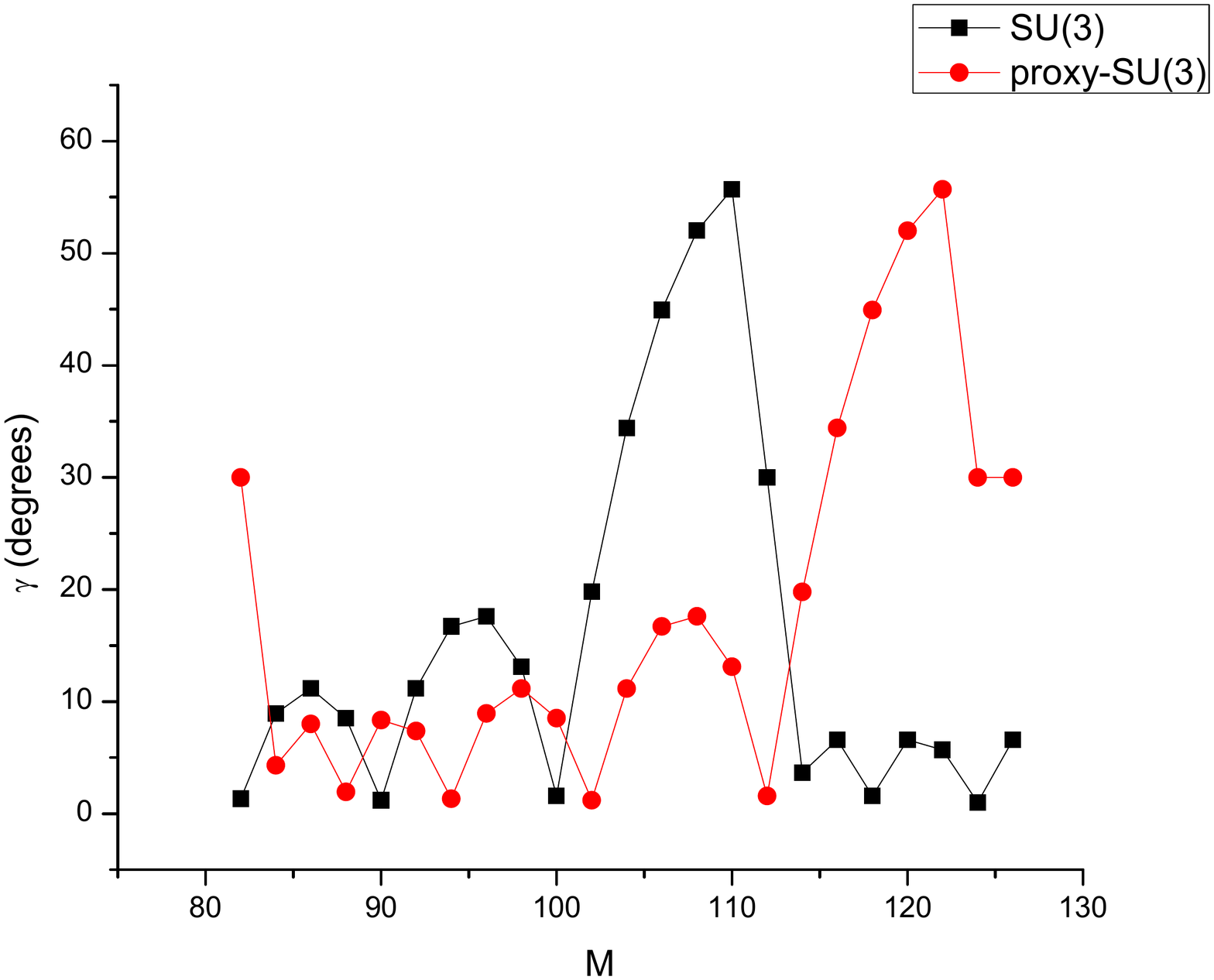}\hspace{1mm}
\includegraphics[width=90mm]{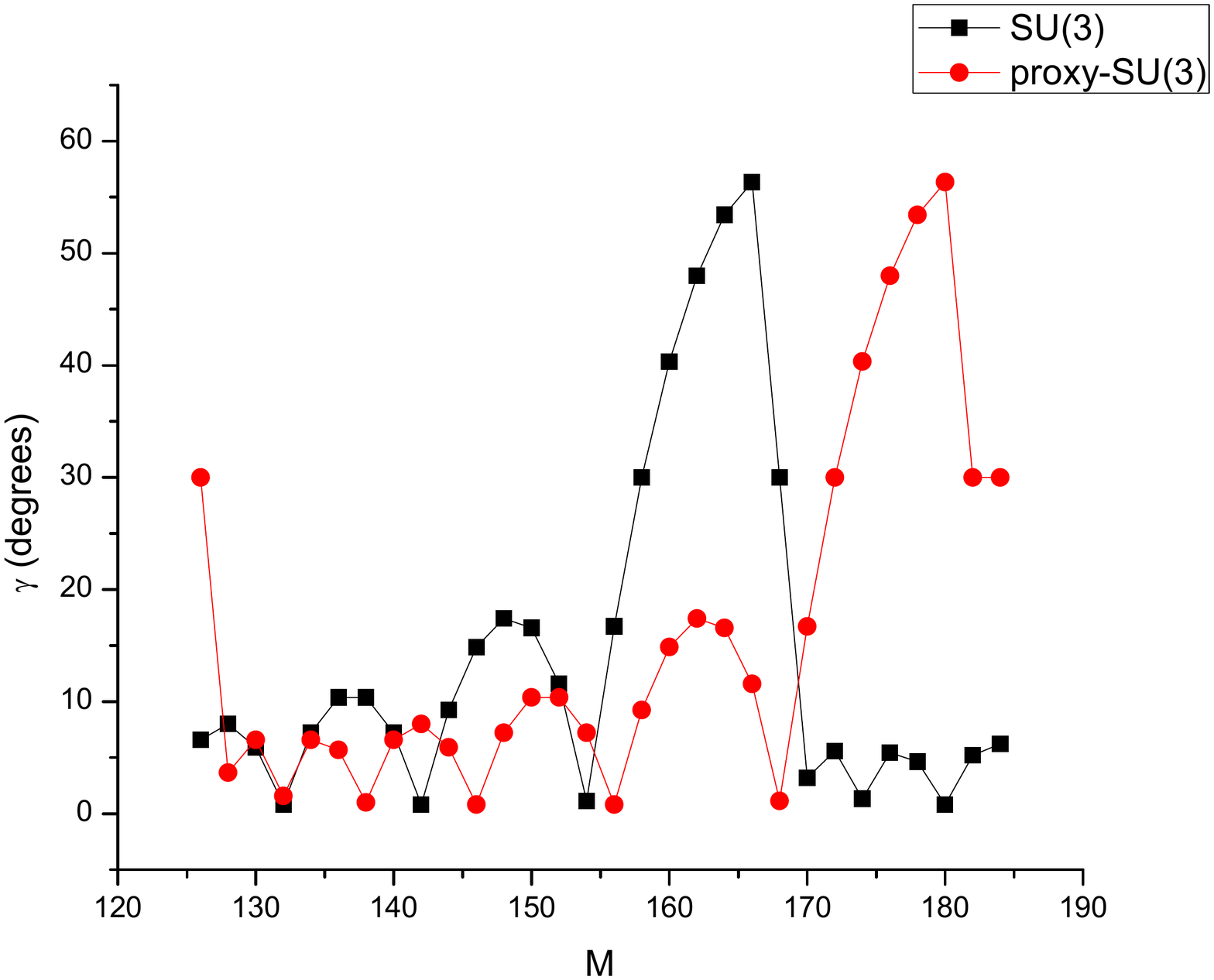}}

\caption{\footnotesize{The deformation parameter $\gamma$ 
(Eq. (\ref{g1})~) \cite{Castanos,Park,proxy2} is plotted against the nucleon (proton or neutron) number $M$ for the 3D harmonic oscillator possessing the SU(3) symmetry, as well as for the proxy-SU(3) scheme. See Section on Shape Coexistence for further discussion.}} 

\end{figure*}
 
\bigskip\noindent
{\bf Regions of coexistence}
\medskip\noindent

In Fig. 1, the quantity $\sqrt{ C_2(\lambda,\mu)+3}$, which is proportional to the deformation variable 
$\beta$ \cite{Castanos,Park} is plotted vs. the nucleon (proton or neutron) number $M$ within the nuclear shells 28-50, 50-82, 82-126, and 126-184, in two different ways, first by using the 3D-HO magic numbers, 
labeled by SU(3), and then by using the proxy-SU(3) magic numbers (which practically coincide with the usual 
nuclear magic numbers produced by the spin-orbit force, taking however into account that the proxy-SU(3) shell can accommodate one pair of nucleons less than the corresponding shell-model shell), labeled by proxy-SU(3). We see that there are specific regions in which the 3D-HO magic numbers lead to lower deformation 
than the spin-orbit magic numbers. These regions are 34-42, 60-72, 96-116, 146-172 respectively. 
However, at this point we should remember that the 3D-HO irreps and the proxy-SU(3) irreps live in the same shell (with the same $N_q$ value) only below the HO magic mumbers 40, 70, 112, 168. Therefore we should take 
into account for further considerations the truncated regions 34-40, 60-70, 96-112, 146-168. This truncation
is necessary in order to keep the two $K^\pi=0^+$ bandheads close in energy, since the addition of an extra oscillator quantum to one of them will make their energy difference very large. 

We remark that the regions 60-70 and 96-112 coincide with the neutron regions in which shape coexistence is observed at $Z\sim 50$ and $Z\sim 82$ respectively. Looking at the map of shape coexistence regions given in Fig. 8 of Ref. \cite{Heyde}, we see that shape coexistence ``bubbles'' do remain within these borders. 

In Fig. 2, the collective variable $\gamma$, calculated through Eq. (\ref{g1}), is plotted vs. the nucleon number $M$. We see that within the regions mentioned above, there are subregions, occupying the right part of the whole regions, in which the 3D-HO magic numbers provide oblate shapes, while the proxy-SU(3) magic numbers give prolate shapes. In particular, these are the subregions 34-40, 64-70, 104-112, 158-168. It is clear that within these regions a prolate band and an oblate band will coexist. This kind of coexistence can be seen as rooted in the particle-hole symmetry breaking mentioned above. 

In addition, in Fig. 2 we see subregions complementary to the ones just mentioned, in which both the 3D-HO magic numbers and the proxy-SU(3) magic numbers provide prolate shapes. These are the subregions 60-62, 96-102, 146-156. In these cases, two prolate bands of different deformation are expected to coexist. 

The above calculation shows that the competition between 3D-HO magic numbers and proxy-SU(3) magic numbers 
can lead to useful observations related to the domains in which shape coexistence can be expected 
across the nuclear chart, leading also to specific predictions about the prolate or oblate character 
of the bands involved. Of course, in order to reach solid conclusions, one has to work out 
the 3D-HO irreps and the proxy-SU(3) irreps for each nucleus separately, and take both protons and neutrons 
into account, as in Ref. \cite{proxy2}. Work in this direction is in progress.

\bigskip\noindent
{\bf Acknowledgements }
\medskip\noindent

Helpful discussions with K. Blaum, R. B. Cakirli, and R. F. Casten are gratefully acknowledged. 
Work partly supported by the Bulgarian National Science Fund (BNSF) under Contract No. DFNI-E02/6.

\end{document}